\documentclass[useAMS,usenatbib]{mn2e}

\usepackage{amssymb}
\usepackage{amsmath}
\usepackage{graphicx}

\newcommand{\f}   {\frac}
\newcommand{\mbfu}{\mathbf{u}}
\newcommand{\ddt}{\partial_t}

\title[Cloud Dispersal in Turbulent Flows]{Cloud Dispersal in Turbulent Flows}
\author[F. Heitsch et al.]{F. Heitsch$^{1,2}$\thanks{E-mail:fheitsch@umich.edu}, 
A.D. Slyz$^{3}$, J.E.G. Devriendt$^{3}$ and A. Burkert$^{2}$\\
$^{1}$Department of Astronomy, University of Michigan, 500 Church St, Ann Arbor, MI 48109-1042, USA\\
$^{2}$University Observatory Munich, Scheinerstr. 1, 81679 Munich, Germany\\
$^{3}$Universit\'e de Lyon 1, Centre de Recherche Astronomique de Lyon, Observatoire de Lyon, 
      9 Avenue Charles Andr\'e,\\
      F-69230 Saint-Genis Laval, France; CNRS, UMR 5574, Ecole Normale
      Sup\'erieure de Lyon}
\begin{document}

\date{Accepted --- Received ---}

\pagerange{\pageref{firstpage}--\pageref{lastpage}} \pubyear{2006}

\maketitle

\label{firstpage}

\begin{abstract}
Cold clouds embedded in warm media are very common objects in astrophysics.
Their disruption timescale depends strongly on the dynamical configuration.
We discuss the evolution of an initially homogeneous cold cloud embedded 
in warm turbulent gas. 
Within a couple of dynamical timescales, the filling factor of the cold
gas within the original cloud radius drops below $50$\%. Turbulent 
diffusivities estimated from the time evolution of radial filling factor
profiles are not constant with time. Cold and warm gas
are bodily transported by turbulence and mixed. This is only mildly indicated
by column density maps. 
The radiation field within the cloud, however, increases by several orders of magnitudes due 
to the mixing, with possible consequences for cloud chemistry and evolution within a 
few dynamical timescales.
\end{abstract}


%
%
\section{Motivation}\label{s:motivation}

Molecular clouds (MCs) in the Galaxy exhibit a wealth of 
structures in (column) densities, velocities and
magnetic fields. The observed structural properties
strongly suggest that MCs are highly dynamical 
objects within a turbulent interstellar medium (ISM). 
Especially, the observed non-thermal linewidths 
(\citealp{FAP1990}; \citealp{WBM2000}) indicate 
the turbulent nature of the clouds. The importance of molecular 
cloud turbulence for the process of star formation
has been the subject of many investigations
(see \citealp{ELS2004} and \citealp{MAK2004} for overviews).

Turbulent mixing is an ever-recurring
theme in the ISM. Classical turbulent mixing
accelerates diffusive transport by a factor
on the order of the Reynolds number of the flow (e.g. \citealp{SCH1977}).
\citet{DAM2002,DAM2003} and \citet{KLL2003} discussed the applicability of
turbulent transport and mixing to a turbulent ISM,
stating that transport rates can vary strongly
with time, sometimes exhibiting super-diffusive behaviour due to
bulk motions of the gas.
Turbulent mixing has also been held responsible as a source
of the highly ionized gas observed around High Velocity
Clouds in the Galactic Halo (e.g. \citealp{FSW2004}),
although it is unclear whether turbulent mixing or evaporation by
heat conduction is the dominating process
\citep{COM1977,MCC1977,BAM1982,LAZ2006}

A spherical cold cloud travelling through a warm uniform
medium dissolves approximately within a timescale 
$\tau_d = M_{cl}/(\rho \pi R_{cl}^2 v)$, where $M_{cl}$ is the cloud's
initial mass, $\rho$ its density, $R_{cl}$ its radius, and
$v$ its velocity \citep{NUL1982}. The main agent is the 
Kelvin-Helmholtz-instability, whose efficiency in dissolving 
the cloud depends on the cloud's physical properties, e.g 
whether it is self-gravitating \citep{MWB1993}
-- in which case there exists a critical mass above which the
cloud remains stable --, or whether it suffers radiative energy 
losses \citep{VFM1997} -- in which case the instability can
be damped, stabilizing the cloud.

Cloud dispersal by hydrodynamical instabilities might be of considerable
importance:
observations and theoretical considerations suggest that molecular 
clouds in the solar neighborhood have lifetimes of approximately
$2$ to $3$ Myrs, which would necessitate close to instantaneous star formation 
once the clouds have formed (\citealp{ELM1993,ELM2000}; \citealp{HBB2001}; 
\citealp{PAL2001}; \citealp{HAR2003}). While stellar feedback in the form of supernovae
could disrupt a cloud, low-mass stars in solar neighborhood clouds 
might not be efficient enough to achieve
such a rapid dispersal (e.g. \citealp{MAC2004}). An alternative
route has been offered by interpreting molecular clouds as
transient objects generated by large-scale colliding flows
(\citealp{BHV1999}; \citealp{HBB2001}; \citealp{HBH2005}; \citealp{VRP2006}; 
\citealp{HSD2006}). In this picture, the flows in which the clouds
form eventually might lead to their dispersal within a few Myrs.

We investigate
the role of turbulence for the overall evolution of 
a cold cloud -- corresponding to the Cold Neutral Medium (CNM) -- embedded in 
warm gas whose parameters are representative of the Warm Ionized Medium (WIM). 
Specifically, we are interested in the timescales on which a 
cold solid cloud will disperse in a turbulent environment, 
and in the evolution of the radiation field within the cloud
(\S\ref{s:physics}). A detailed, time-independent study of the radiation
field in a filamentary cloud can be found in \citet{BZH2004} and \citet{BZL2006}.

We find (\S\ref{s:results}) that the warm and cold components
are efficiently mixed by bodily transport. However, the central 
optical depth stays comfortably above $1$ even for late times. At first,
this might seem contradictory, however, the
local radiation field within the cloud increases by several
orders of magnitude within a couple of dynamical times.
Turbulent diffusivities derived from the expansion of the cold cloud generally
are not constant with time. Possible consequences for cloud lifetimes 
and evolution are discussed in \S\ref{s:summary}.

%
%
\section{The Problem and its Setup}\label{s:physics}

We start with a uniform sphere of cold dense gas in a warm diffuse 
ambient medium. The system is initially completely in thermal and almost in 
turbulent pressure\footnote{As discussed by e.g. \citet{BAL2006}, 
the concept of turbulent pressure is only applicable if there exists a
scale separation between the (large) object scale and the (small)
turbulent scale. In a medium without scale separation -- e.g. with a
self-similar turbulent power spectrum --  the term ``turbulent pressure'' 
should be read as ``turbulent kinetic energy density''. It is in 
this sense we will use the term ``turbulent pressure'' throughout the paper.}
equilibrium. One model
set (sequence A) is purely adiabatic, i.e. the two-phase structure
of the system will get erased with time as gas with temperature intermediate
to the two initial temperatures is created. The second set (sequence C)
keeps the identities of the warm and cold phases distinct by a combination of heating
and cooling processes \citep{WHM1995} typical for the warm and cold
atomic ISM. Thus, we study turbulent mixing under conditions with and without
radiative losses. Initially, both gas components are in thermal equilibrium at temperatures
that for the radiative case correspond to the two stable temperature regimes, i.e. 
there is no gas in the thermally unstable regime
initially. The lower initial temperature is set to $T=31$K, and
the density contrast between warm and cold gas is $300$, with $n_0 = 0.5$cm$^{-3}$ in the 
warm phase at $T=9.2\times 10^3$K.
The cubic box is periodic in all directions, with a side length, $L$, 
of $44$pc, and the cold gas sphere starts out with a radius of $4.4$pc. 

Instead of studying the cloud evolution within a shear flow,
we resort to a more direct way of treating turbulence, namely
by initially imposing a velocity field drawn from a random Gaussian distribution
(see e.g. \citealp{MKB1998}). Power is alotted in Fourier space only to the largest
scales, and with random phases. This is meant to mimic the effect of  
turbulence generated by an unspecified source on larger scales.
The initial Mach number in the warm gas ${\cal M}_0=2$ or $3$
-- corresponding to $v_0 = 22.4$km s$^{-1}$ and $v_0=33.6$km s$^{-1}$ -- 
is higher than Mach numbers commonly
ascribed to the warm ISM, however, since the turbulence is not driven,
the system acquires reasonable values of ${\cal M}$ once turbulence
is fully developed. The justification for this will be discussed in \S\ref{ss:turbdiss}.
One might argue that the scenario of a uniform, spherical cold cloud evolving in a decaying 
turbulent velocity field  is only of limited physical relevance, since
the Galactic ISM turbulence is to a large extent thought to be driven by supernovae 
\citep{COS1974,MCO1977,ROB1995,DAV2000,DAB2001}. However, the goal of this study
is not to model the evolution of a cold (possibly molecular) cloud, but to demonstrate that even
under unfavorable conditions (decaying turbulence, no stellar feedback, no gravitational
fragmentation, starting with a uniform spherical cloud), 
cold clouds fragment and disperse sufficiently within a dynamical
timescale so that the internal structure of the cloud is thoroughly altered.
Driven turbulence and/or stellar feedback obviously would lead to faster dispersal, while
gravitational fragmentation would result in a smaller gas filling factor, thus opening holes
and channels for radiation and energy (possibly in form of waves, see e.g. \citealp{HEI2006}) 
to enter the cloud. For this reason also a non-uniform cloud would be more prone to 
disperse in a turbulent environment.

The adiabatic model sequqence 
we will denote by A, the radiatively cooled model by C. 
The initial Mach number of the model is denoted by ``2'' or ``3'', for Mach $2$ or $3$.
We ran models at linear resolutions of $128^3$ and $256^3$ cells, 
indicated by the letters $a$ and $b$ respectively. 
The choice of the -- rather small --
initial cloud radius could raise concerns about how well even the $256^3$-models
are resolved. Thus, we ran an additional model, A2l, with the same parameters
as the adiabatic model at Mach $2$, A2b, except for the initial cloud radius, 
here set to $8.8$pc.

Initially, the system is in thermal and near turbulent pressure balance, 
i.e for turbulent pressure balance
\begin{equation}
  \langle{v^2}\rangle_c = \f{\rho_w}{\rho_c}\langle{v^2}\rangle_w\label{e:initpress}
\end{equation}
holds, with velocities $v$ and densities $\rho$. The indices stand for the cold
and warm phase. The near pressure balance reduces motions due to
pressure differences and therefore allows us to make more valid 
statements about turbulent transport or mixing in this idealized setup. 

The numerical scheme is based on the 2nd order Bhatnagar-Gross-Krook formalism 
(\citealp{PRX1993}; \citealp{SLP1999}; \citealp{HZS2004}; \citealp{SDB2005}),
allowing control of viscosity and heat conduction. The code evolves the
Navier-Stokes equations in their conservative form to second order in time and
space. The hydrodynamical quantities are updated in time unsplit form.

We employed the same heating and cooling prescriptions as \citet{HBH2005,HSD2006}, 
based on \citet{WHM1995}. The same caveats apply, especially that
we are discussing the mixing between the warm and cold ISM, and that we are 
not including molecular gas. Thus, while the parameters of our cold cloud
are consistent with values for ``Giant Molecular Clouds'', we neglect the 
effects of molecular line cooling and chemistry.

The code is equipped with Lagrangian tracer particles that are initially deployed 
within the cold cloud at a resolution of one particle per grid cell. The particles 
are advected with the gas flow, so that they allow us to study the history of 
the cold gas.

We restricted the models to hydrodynamics with heating and cooling, leaving
out gravity and magnetic fields. Depending on their strength, fields could
suppress shear instabilities, while
gravity might lead to more compact dense structures and fragmentation. 
This could have a twofold effect, as will be discussed in \S\ref{s:summary}.

%
%
\section{Results}\label{s:results}

\subsection{Morphologies}\label{ss:morph}

\begin{figure}
  \includegraphics[width=\columnwidth]{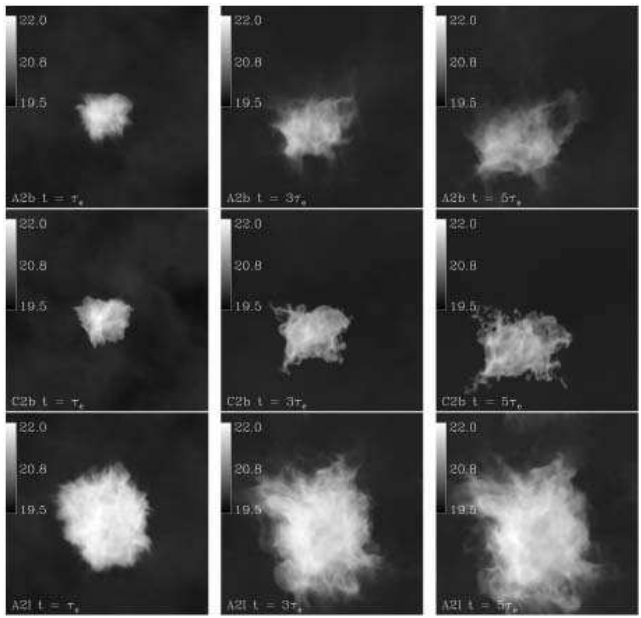}
  \caption{Column density maps of model A2b (top), C2b (center), 
           and A2l (bottom) projected along the $z$-axis, 
           for times $t=\tau_e$, $3\tau_e$ 
           and $5\tau_e$, where $\tau_e\approx 2$Myrs is the initial turbulent 
           crossing time. Color corresponds to $\log N [\mbox{cm}^{-2}]$.
           \label{f:morphall}}
\end{figure}

A first impression of the efficiency of turbulent mixing 
can be gleaned from a time sequence of column density maps 
(Fig.~\ref{f:morphall}). Column densities
are integrated along the $z$-axis 
and shown at times $t=\tau_e$, $t=3\tau_e$ and $t=5\tau_e$,
where $\tau_e = L/v_0\approx 2$Myrs, a nominal turbulent flow crossing time
in terms of the total box length and the rms velocity in the warm diffuse
gas. Despite the fact that we are looking at column densities here, the
overall effect of the cloud's turbulent dispersal is clearly visible.

The most noticeable difference between the adiabatic models A2b and A2l 
(top and bottom) on the one hand and the radiative
models (C2b, center) on the other is that for the latter, the 
transition between low and high column
densities is much more marked, i.e. the column density maps appear (especially
in the later stages) less ``frothy''. As we will see below, this is a direct
consequence of the cooling.

In overall appearance, for both the adiabatic and radiative models 
the cloud stays more or less compact, although
fingers of cold, dense material are clearly sprouting from the core.

\subsection{Mass fractions and Dynamics of the Warm and Cold Gas}\label{ss:dynphas}

\begin{figure}
  \includegraphics[width=\columnwidth]{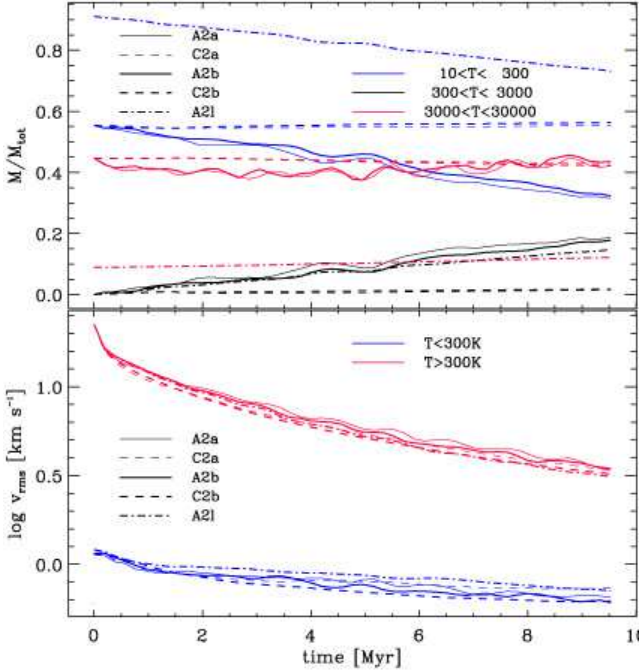}
  \caption{{\em Top}: Mass fraction over the whole simulation volume
           for the three thermal regimes indicated, against
           time, for models A2a/b and C2a/b at resolutions $N=128^3$ and $N=256^3$, and
           for model A2l (at $N=256^3$).
           {\em Bottom: }Rms velocity $\langle{v^2}\rangle^{1/2}$ for cold and warm gas
           against time, for the same models and resolutions. Note that red and blue
           lines here denote $T>300$K and $T<300$K respectively.
           \label{f:bulkmeas}}
\end{figure}

Since we will be discussing the turbulent mixing of the warm and
cold gas, first, we need to understand the dynamics and 
evolution of each of the phases. Note that -- strictly speaking --
the concept of ``phases'' can be misleading not only for the
adiabatic case, but also for the thermally bistable case,
because of the importance
of dynamics (see e.g. \citealp{VGS2000}; \citealp{VRP2006}; \citealp{HSD2006}).
Motivated by the bistable models (sequence C), we split the 
model-ISM into three regimes, namely a cold phase with temperatures
$300\mbox{K} > T$, a warm phase ($3000\mbox{K} < T$) and an 
intermediate phase with $300\mbox{K}<T<3000\mbox{K}$. 
The corresponding mass fractions (Fig.~\ref{f:bulkmeas}, top panel) evolve
quite differently for adiabatic and radiative models. The mass fractions
are taken over the whole simulation volume. Since the cloud in model
A2l has an initial radius twice as large as in the other models,
it starts off with a larger cold mass
fraction and a smaller warm mass fraction. The intermediate 
temperature regime evolves similarly to that in models A2a/b. 

In the adiabatic case (A2a/b/l), some of the cold gas is lost to the 
intermediate regime, while the mass fraction in the warm phase 
stays pretty much constant.
In contrast, the radiative case (C2a/b) keeps constant mass fractions
in each of the three temperature regimes over the whole simulation time 
(i.e. $5$ dynamical times). Heating and
cooling timescales are much shorter than the dynamical timescales,
so that gas cannot collect in the intermediate regime, which for
model C2a/b corresponds to the thermally unstable regime. In particular,
in C2a/b gas which is heated by (viscous) shear at the cloud boundaries 
 ``falls back'' to its previous (cold) thermal state, while 
for A2a/b, this gas has no way to return to its previous temperature
except by adiabatic cooling. Varying the resolution does not change
the mass fractions appreciably. Model C2a/b displays a tiny increase
of the cold mass fraction. This stems from the compression of warm
gas when it collides with the cold cloud rim. Once the warm gas' density
increases, cooling sets in, and this compressed gas is added to the cold
gas component. 
However, as Figure~\ref{f:bulkmeas} shows, this effect is negligible.

In contrast to the evolution of the mass fractions differing for the adiabatic and
radiative models, the rms velocities of the gas in the cold and warm temperature regimes
evolve similarly for both sets of models.
Figure~\ref{f:bulkmeas} (bottom panel) mirrors the initial pressure balance 
(eq.~\ref{e:initpress}): velocities in the cold phase start out lower by 
a factor of approximately $17$. For each of the models A and C, the velocities
in the cold and warm phase decay, albeit at different rates. Resolution effects
do not affect the decay (the thin and thick lines are nearly indistinguishable). 
Comparing model A to C, the radiative losses occurring in C do not lead to
significantly different decay rates. 
Radiative losses would become important in regions
of high compression, however, the turbulence initially decays quickly below
Mach $1$ in both the warm and cold gas, limiting the compression. 
We discuss the time evolution of the cold and warm pressure profiles in detail
in \S\ref{ss:radprof}. The larger cloud (model A2l) shows the same velocity
decay as its small counterparts.

\subsection{Radial Profiles}\label{ss:radprof}

\begin{figure}
  \includegraphics[width=\columnwidth]{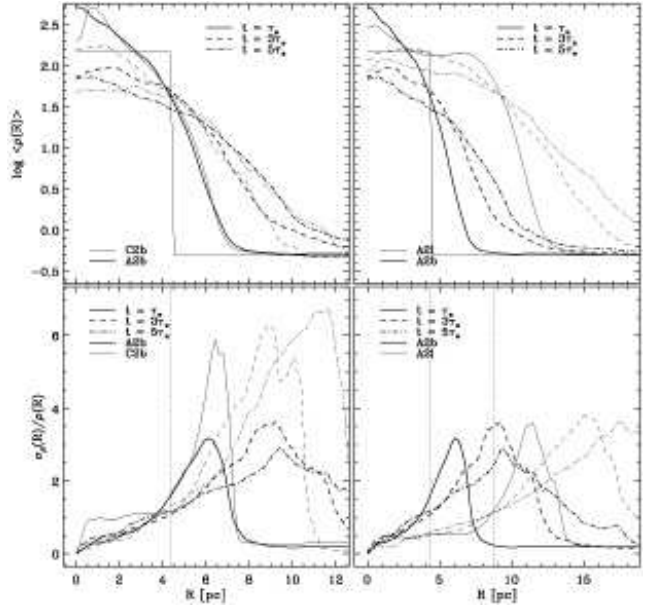}
  \caption{{\em Top:} Radial density profiles for models A2b, C2b and A2l,
           at the times indicated.
           Model A2b is shown in thick lines in both plots. 
           The thin solid step function gives the initial density profile for
           models A2b and C2b. The initial density profile of model A2l
           extends out to $8.8$pc.
           {\em Bottom:} Relative standard deviation of density profile 
           (eq.~\ref{e:relstddev}). The dotted vertical line denotes the initial radius.
           \label{f:rhorad}}
\end{figure}

Does the cloud break up in the turbulent environment, and 
if so, how quickly does this happen? The answer depends
strongly on the quantity we are looking at. Since we are
interested in an average measure of the cloud's structure,
we take averages over shells and discuss the resulting 
radial profiles. The shell centers coincide with the
instantaneous center of mass of the cloud (the cloud itself moves
a little in the background flow).
We begin with the density profiles (Fig.~\ref{f:rhorad}). 

Clearly, the turbulent motions lead to a spreading-out of the cloud, mirrored
in the density-weighted average radius 
\begin{equation}
  \langle{R}\rangle_M\equiv \f{\int_0^{R}r\rho(r)dr}{\int_0^{R}\rho(r)dr}\label{e:avgrad}
\end{equation}
of the cloud, which roughly doubles within $5\tau_e = 10$Myrs.
As Figure~\ref{f:morphall} already indicates, the radial density
profiles show a substantial variation at fixed radius $R$ (Fig~\ref{f:rhorad}, lower
panel), up to six times the actual density value. The quantity 
\begin{equation}
  \sigma_\rho(R)/\rho(R)\equiv\left\langle(\rho(R)-\langle\rho\rangle)^2\right\rangle^{1/2}\,/\,\rho(R)
  \label{e:relstddev}
\end{equation}
plotted is the relative standard deviation on the mean density over a sphere at radius $R$.
The strongest variations are
expected at the cloud rim, which travels (upper panel) outwards, so that the peak
of the density variations is seen at larger radii for later times. 
Already after one dynamical time, the cloud is far from being a solid sphere.

\begin{figure}
  \includegraphics[width=\columnwidth]{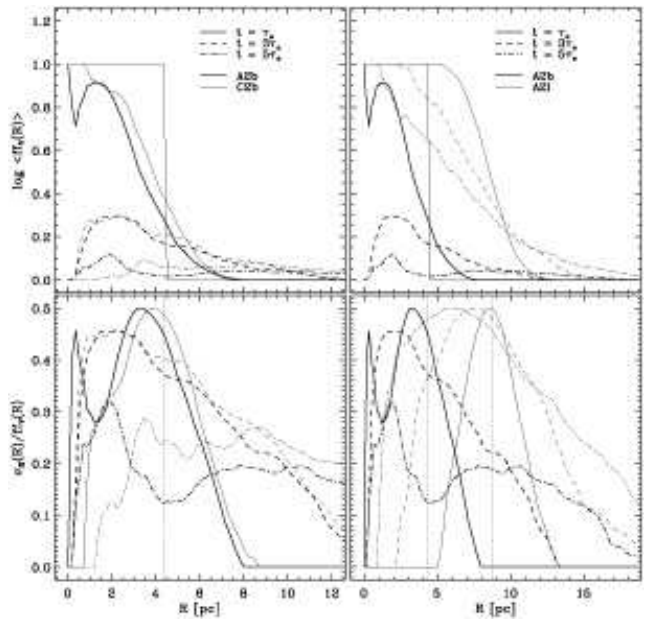}
  \caption{{\em Top:} Radial cold gas filling factor profile for models A2b, C2b and A2l
           at the times indicated.
           Model A2b is shown in thick lines in both plots. 
           The thin solid step function gives the initial cold gas filling factor profile for
           models A2b and C2b. The initial profile of model A2l
           extends out to $8.8$pc.
           {\em Bottom:} Relative standard deviation of filling factor profile 
           (analogous to eq.~\ref{e:relstddev}). The dotted vertical line denotes the initial radius.
           \label{f:filfacrad}}
\end{figure}

A more stringent measure is the radial volume filling factor profile for gas in the cold
phase (Fig.~\ref{f:filfacrad}). Volume filling factors are measured on shells at given $R$.
As in \S\ref{ss:dynphas}, the temperature threshold to
distinguish between the warm and cold phase is set at $T=300$K.
The step function in the upper panel gives the initial condition, which of course
shows a cold gas filling factor of $1$ for $R\leq 4.4$~pc (or $R\leq 8.8$~pc for model A2l). 

The first obvious difference between the radiative and adiabatic case is that
the adiabatic case seems to mix faster the cold and 
warm phase at more radii. However, this is not that surprising, since for the adiabatic case
{\em any} gas with $T>300$K is lost for the cold gas filling factor, while for the 
radiative case, gas which has left the cold phase can only be found in the warm 
phase (there is no intermediate-phase gas, Fig.~\ref{f:bulkmeas}, top panel) but
because of the short cooling times, this gas can quickly return to the cold phase.
In other words the filling factor gives an unambiguous measure of the degree of mixing
between cold and warm gas only for the thermally bistable, radiative case (C2b),
since there is (close to) zero conversion between the gas phases (see also discussion
on tracer particles below).

\begin{figure}
  \includegraphics[width=\columnwidth]{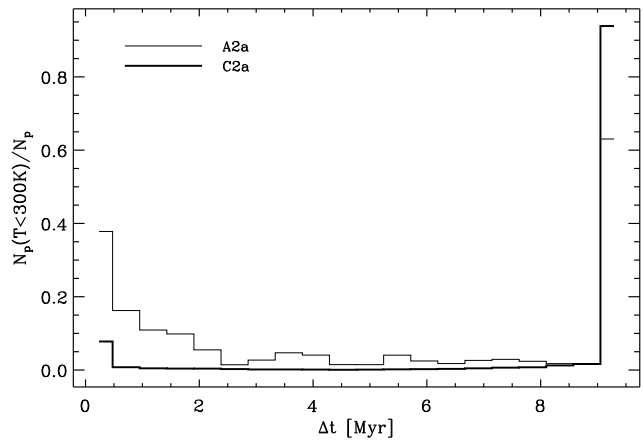}
  \caption{Histogram of fraction of tracer particles in the flow staying at $T<300$K
           for a time interval $\Delta t$. 
           Most tracer particles stay cold over the whole simulation time 
           ($\Delta t>9$Myr) for model C2b (thick line), while for model A2b
           (thin line), only $\sim 60$\% of the particles stay at $T<300$K.
           \label{f:temphist}}
\end{figure}

After $2$ Myrs (corresponding to $t=\tau_e$), the volume filling factor
of the cold gas measures $40$\% for the radiative
case, and $\sim 25$\% for the adiabatic case at the initial cloud boundary, 
i.e. more than half of the volume is occupied by warm gas. 
Note that especially in the radiative 
case, the {\em mass fractions} of cold and warm gas stay constant, i.e. cold and
warm gas are bodily transported. This can be gleaned from Figure~\ref{f:temphist}.
It shows the histogram of the fraction of Lagrangian tracer particles $N_P(T<300\mbox{K})/N_P$ 
within the cloud that stay at temperatures $T<300$K for a time interval 
$\Delta t$. If there were no turbulence in the 
models, all the particles would stay at $T<300$K for the whole duration of the 
simulation, i.e. we would have $N_P(T<300\mbox{K})/N_P=1$ at $\Delta t = 10$.

Most of the particles stay cold for longer than $9$Myr, 
i.e. for the  simulation's time extent. In other words, the constancy of the mass fractions
in Figure~\ref{f:bulkmeas} indeed results only to a very minor extent from 
the conversion of cold to warm gas and
vice versa: mass in the cold and warm regime is separately conserved. This is less
valid for the adiabatic case, which obviously loses some of its cold material.
But still, the bulk of the initially cold gas stays cold for the whole simulation
time. 

\begin{figure}
  \includegraphics[width=\columnwidth]{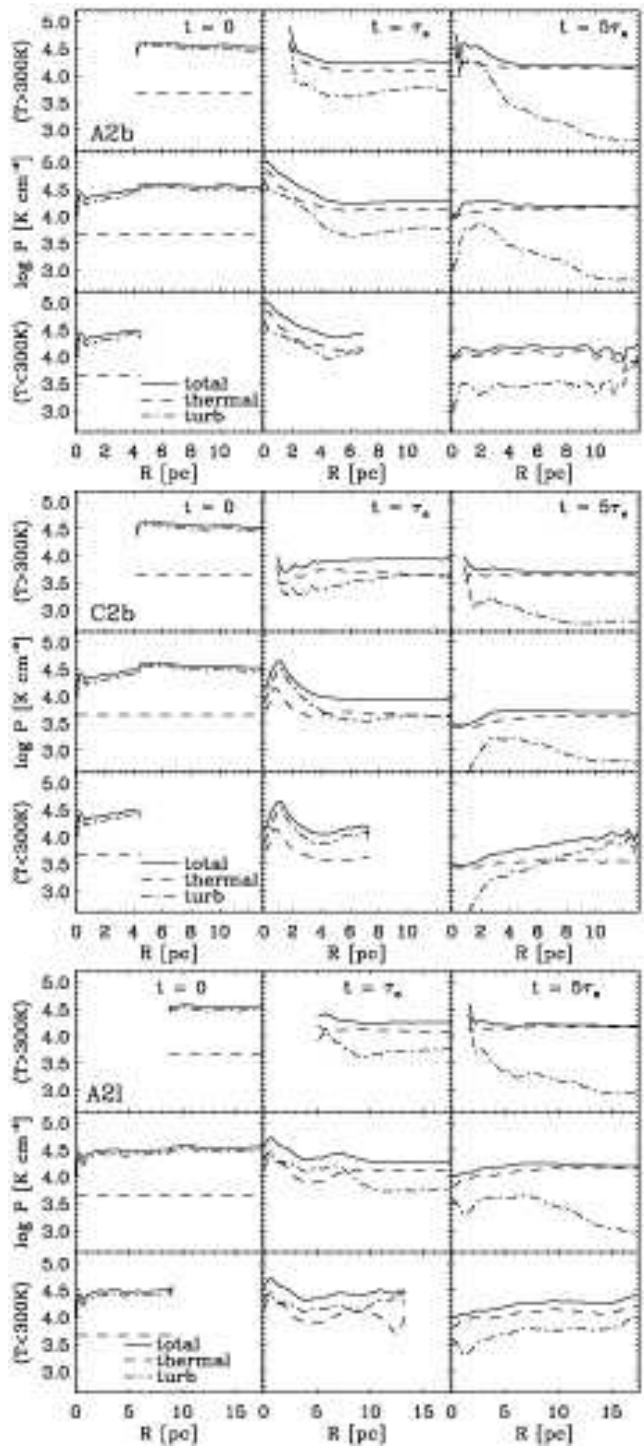}
  \caption{Radial pressure profiles for models A2b (top), C2b (center) and 
           A2l (bottom) for
           times $0$, $\tau_e$ and $5\tau_e$. In each panel, from top to bottom:
           radial pressure profile for gas at $T>300$K, total radial pressure
           profile, and radial pressure profile for gas at $T<300$K. 
           Plotted are the total pressure (solid lines), the thermal pressure
           (dashed) and the turbulent pressure (see text, dash-dot). The
           three columns per model denote the measurement times $t=0$,
           $t=\tau_e$ and $t=5\tau_e$.
           \label{f:radpress}}
\end{figure}

Besides turbulent transport, an overpressure in the cold cloud could lead to ``expansion''.
So far, our diagnostics cannot distinguish between these two mechanisms for cloud
expansion. Figure~\ref{f:radpress}
shows the radial profiles of the pressures for models A2b, C2b and A2l. 
The total pressure $P_{tot}$ has been split into the thermal pressure $P$,  
and the turbulent pressure $P_{trb}$, from which we have removed the 
contribution of the translational velocity $\langle v\rangle$, as
\begin{equation}
  P_{tot} = P + \rho(v-\langle v\rangle)^2,
\end{equation}
where the average extends over coherent cold and warm regions. 
The top row gives the radial pressure profile for gas with $T>300$K, and 
the bottom row shows the profile for gas with $T<300$K. The time 
sequence reveals the mixing of the warm
and cold component, since $P(T>300\mbox{K})$ can be defined at smaller radii for later times,
while $P(T<300\mbox{K})$ spreads outward to large radii resulting in warm and cold
gas co-existing at an increasingly larger radial range. For $t=0$, models A2b and C2b are of 
course identical. The thermal pressure is constant, while the turbulent pressure (and thus
the total pressure) is slightly lower within the cold cloud than in the warm medium.
This serves as a safeguard against initially overpressuring the cloud by 
turbulent pressure. At $t=\tau_e$ (center column), the cold dense 
material (i.e. the cloud) in both models is overpressured relative to the warm gas.
 This comes mostly from an overshoot in the density, since turbulent pressure and thermal 
pressure have (approximately) the same radial dependence: the cloud is initially 
slightly compressed by the higher turbulence in the warm medium (see also the 
radial density profiles, Fig.~\ref{f:rhorad}). In Figure~\ref{f:bulkmeas} 
(bottom panel) we saw that the rms velocity -- and because of the close 
to constant density in the respective phases, the kinetic energy -- 
decays faster for the warm gas than for the cold gas.
Both effects together lead to the overpressure in the cold gas at
$t=\tau_e$. At later times, this pressure imbalance has canceled out, and the
warm and cold phases are mixed. There is still a turbulent pressure excess
at small radii, however, the turbulence has decayed so far that its dynamical effects
are insignificant. 

For the adiabatic run, at $t=\tau_e$ the thermal pressure 
has risen from its initial value both for the cold and warm gas. The turbulent 
pressure on the other hand, has dropped by about a factor $3$ for the warm gas, 
but not as much for the cold gas, suggesting that turbulent energy
in the warm phase has been used to heat the warm phase and to
drive the turbulence in the cold phase, since the thermal energy in the cold phase 
increased and the turbulent energy remains unchanged. With the total energy conserved,
some of the energy in the cold gas has to come from the warm gas.
For the radiative model at  $t = \tau_e$ the transfer of energy from the warm to the cold gas 
is less marked but seems to have occurred nevertheless. The cold gas' thermal and turbulent 
pressure in the innermost radii increase from their initial values. The thermal pressure of the
warm gas on the other hand has hardly changed from its initial value even though the turbulent
pressure has dropped by almost an order of magnitude from its initial value at all radii
where warm gas exists. Again, the thermal and/or turbulent pressure increase in the cold gas
has to come from the warm gas. Model A2l shows essentially the same behavior as model A2b.

\begin{figure}
  \includegraphics[width=\columnwidth]{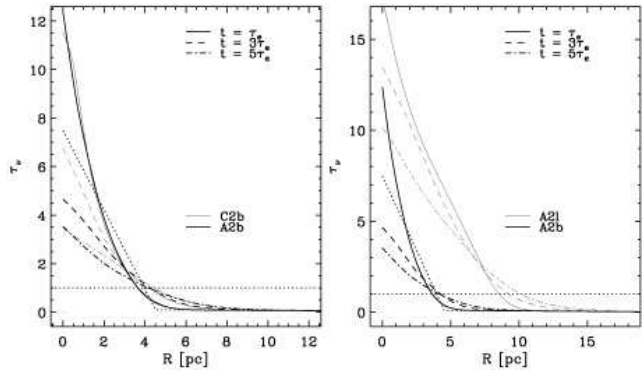}
  \caption{{\em Left:} Radial optical depth profile for models C2b (thin lines) and
           A2b (thick lines), at the times as indicated in the plot.
           The thick dotted line gives the initial optical depth profile, while
           the thin dotted line denotes $\tau_\nu = 1$. {\em Right:} Same plot
           for models A2l (thin lines) and A2b (thick lines). Note that the
           optical depth range is larger for model A2l.
           \label{f:optdeprad}}
\end{figure}

The radial density profile allows us to determine the average optical depth at a given
radius (Fig.~\ref{f:optdeprad}). To arrive at meaningful optical depths, we scaled
the effective absorption coefficient $\alpha_\nu$ such that the central optical depth is 
initially arbitrarily set to $\tau_c = 7.5$ for models A2a/b and C2a/b, 
and to $\tau_c=15$ for model A2l. This corresponds to an absorption coefficient of 
$\alpha_\nu=3.6\times 10^{-21}$ cm$^2$. For comparison, Spitzer (1961) gives an 
effective combined cross section for absorption and scattering on grains of 
$\alpha_\nu = 1.2\times10^{-21}$ cm$^2$ at a wavelength of $100$ nm. Likewise, we do not
take into account the effect of scattered light.

There are no marked differences between the adiabatic and the radiative case. 
This as well as the fact that $R(\tau=1)$ does not change appreciably over time
is a direct consequence of the radial density profiles (Fig.~\ref{f:rhorad}).
At $t=\tau_e$, the optical depth in the center has increased due to an initial
compression, just to drop at later times because of turbulent dispersal. 
This ``overshoot'' is mirrored in the density profiles (Fig.~\ref{f:rhorad}, top panel). 

The central optical depth $\tau_c$ drops by approximately a factor of $2$, but does not
fall below $\tau = 1$. That $\tau_c$ changes at all with time might come as a surprise,
but is a consequence of the exchange of dense and diffuse material at approximate
pressure equilibrium: the filling factor of dense material on a spherical surface at
fixed radius decreases, lowering the central optical depth.

\subsection{Turbulent Diffusivity}\label{ss:turbdiss}

The evolving radial density and filling factor profiles 
(Figs.~\ref{f:rhorad} and \ref{f:filfacrad}) suggest to
estimate the (turbulent) diffusivity by fitting them to 
the expected profiles resulting from the time evolution
of a step function under the effect of diffusion.
An (inert) quantity $q$ in a turbulent environment obeys the advection-diffusion equation
\begin{equation}
  (\ddt + \mbfu\cdot\nabla)q = \lambda\nabla^2q,\label{e:advediff}
\end{equation}
where $\mbfu$ is the velocity, and $\lambda$ the microscopic diffusivity, which
has been assumed to be independent of location and direction. Equation~\ref{e:advediff}
can be rewritten (see e.g. appendix in \citealp{HZS2004}) as 
\begin{equation}
  \ddt\langle q\rangle = (\lambda_e+\lambda)\nabla^2\langle q\rangle,\label{e:qldtdiff}
\end{equation}
under the -- contestable -- assumption of a separation between the small scale turbulent
velocity field and the large scale variations in the quantity $\langle q\rangle$, where
the averages have removed variations due to the small-scale turbulence. The ``turbulent
diffusivity'' $\lambda_e\equiv u_{rms}L$ is the product of the rms velocity and the 
characteristic length scale over which a gas parcel maintains $u_{rms}$ (see e.g. \citealp{LAL1966}).
In the ISM, the Reynolds number $Re\equiv\lambda_e/\lambda\gg 1$, generally, so that $\lambda$
can be neglected for turbulent transport studies. Quasi-linear diffusion theory 
(see e.g. \citealp{MOF1978}) holds that $\lambda_e$ can be regarded as constant.

Since our model clouds start out with a uniform density, we can follow the discussion by
\citet{DAM2002} and study the turbulent diffusive evolution of a step function profile. 
A one-dimensional density distribution evolves as 
\begin{equation}
  n(x,t) = \f{1}{2\sqrt{\pi\lambda_e t}}\int^{\infty}_{-\infty}n(x',t=0)\,e^{-(x-x')^2/4\lambda_e t}dx'
  \label{e:profile}
\end{equation}
under diffusion. The initial conditions can be written as $n(x\leq R_0,t=0) = n_1$ and 
$n(x > R_0,t=0) = n_0$, where $R_0$ is the initial cloud radius at $t=0$. 
Then, the density distribution at time $t$ is given by
\begin{equation}
  n(x,t) = n_0+\f{n_1-n_0}{2}\left(1-\mbox{erf}(\f{1}{2}\f{x-R_0}{\sqrt{\lambda_e\,t}})\right),\label{e:diffdens}
\end{equation}
where $\mbox{erf}(x)$ is the error function.
A similar expression is valid for the filling factor $f(x,t)$, if we replace $n(x,t)$ by 
$f(x,t)$, and set $n_0=0$ and $n_1=1$.  
Since the initial radius $R_0$ is known, we can fit equation~\ref{e:diffdens} to the
available density and filling factor profiles at given times $t$, and thus determine the
diffusivity $\lambda_e$ by a 1-parameter Levenberg-Marquardt least-squares minimization
(e.g. \citealp{PTV1992}) for each available time step of a model.

\begin{figure*}
  \includegraphics[width=\textwidth]{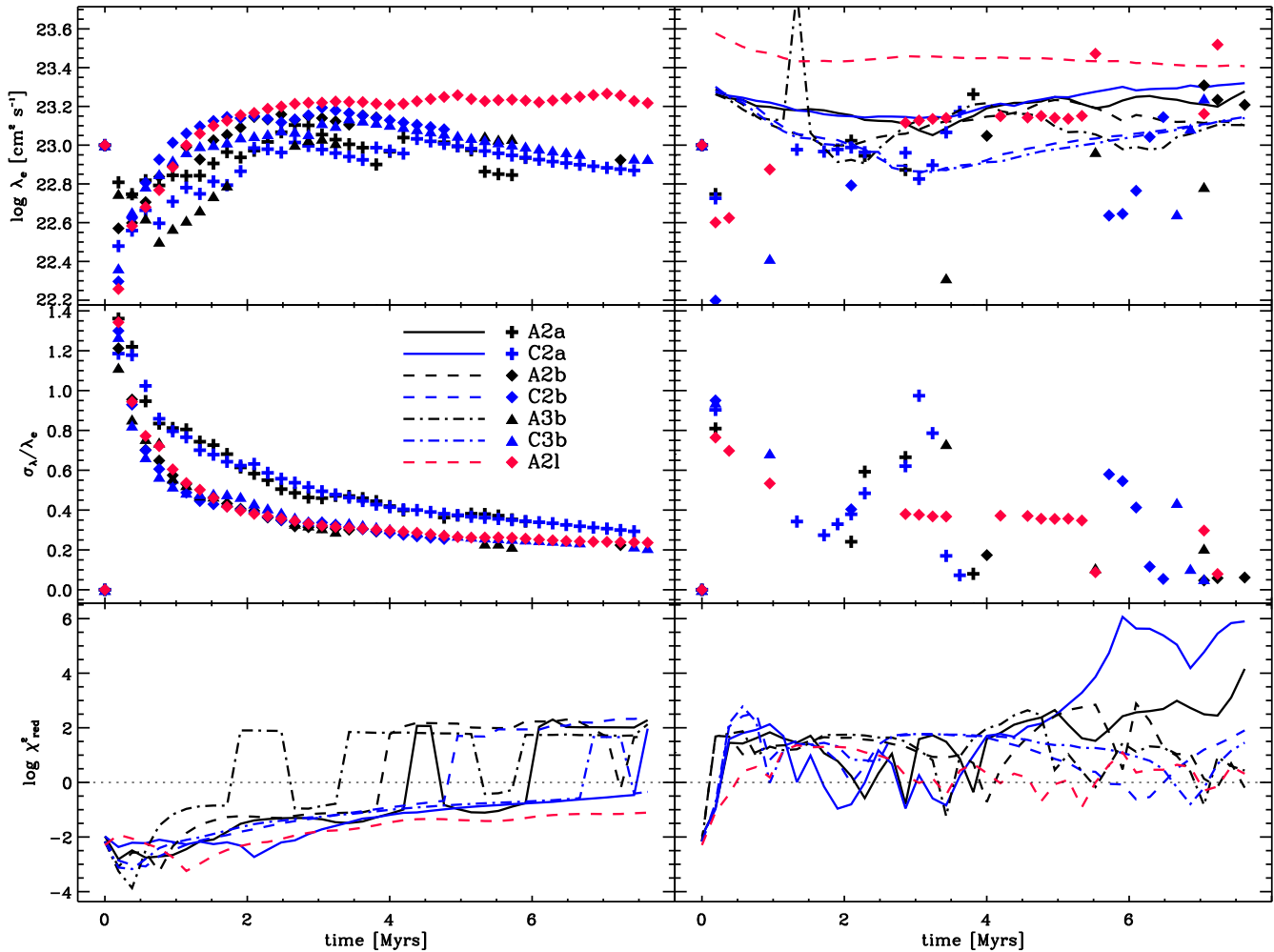}
  \caption{Diffusivities (top row), 1-$\sigma$ errors for profile fits (center row),
           and reduced $\chi^2$ for profile fits (bottom row), against time for all models.
           Left panels show results derived from filling factor profiles and right panels
           show results derived from density profiles.
           Symbols in the top row panels denote diffusivities derived from profile fits 
           (eq.~\ref{e:diffdens}), lines in the top right hand panel
           stand for $\lambda_e$ derived from eq.~\ref{e:diffrms}. Only fit diffusivities 
           for $\chi^2_{red}<2$ are shown.\label{f:difftime}}
\end{figure*}

Figure~\ref{f:difftime} shows the resulting diffusivities for the filling factor profiles
(top left) and the density profiles (top right) for all models. The center row gives
the $1$-$\sigma$ errors of the profile fits, and the bottom row contains the reduced
$\chi_{red}^2$ for the filling factor profiles (bottom left) and for the density 
profiles (bottom right). Reliable fits we select (somewhat arbitrarily) by 
 $\chi_{red}^2<2$, agreeing well with a selection by eye.

Clearly, the filling factor profiles lead to much better fits than the density profiles.  
Figures~\ref{f:rhorad} and \ref{f:filfacrad} explain this:
the filling factor cannot rise above $1$ or drop below $0$, thus
constraining the profiles for the least-square minimization, while the density profiles
actually increase above $n_1$ and drop below $n_0$ because of traveling waves. 
Moreover, in deriving 
equation~\ref{e:diffdens}, we assumed that $n(x\leq R_0,t=0)=n_1$ is valid for all $x<R_0$,
and not only for $0\leq x \leq R_0$, as by construction of the initial conditions.
This assumption is certainly not valid any more at later times.

We first notice that the diffusivities determined by least-square fitting 
(symbols in top row of Fig.~\ref{f:difftime}) are not constant
with time. Until $\sim 3$Myrs they all increase. The increase however is less
than an order of magnitude. The filling factor profiles return slightly decreasing 
diffusivities for later times and for all models except model A2l. \citet{DAM2002} 
found that the diffusivities increase exponentially 
with time. Since they start with an unperturbed medium and then drive the turbulence
via supernova explosions, the growing diffusivities could be a result of the increasing
rms velocity. In our models, a similar effect is causing the initial rise of 
$\lambda_e$. The turbulent diffusivity increases while the turbulent cascade is 
building up. At later times, the overall decay of the turbulent rms velocity 
then leads to a decrease of the turbulent diffusivity.

The density profiles do not provide as nearly a complete picture as the filling factor
profiles. Reliable fits are difficult to get because of shock waves traveling through
the cloud. The few reliable fits clearly show that the diffusivities are not constant,
but range within the same values as the ones derived from the filling factor profiles.
As a crude check, the classically expected diffusivities 
\begin{equation}
  \lambda_e\equiv\langle R\rangle_M\,u_{rms}\label{e:diffrms}
\end{equation}
are shown in the top right panel of Figure~\ref{f:difftime} in lines. Initially,
they decrease slightly, mirroring the decay of $u_{rms}$, while the growing
cloud radius compensates for the decaying velocity at later times. 
The values are within a factor of $2$ at most times with the
diffusivities derived from the profile fits. A $\lambda_e=3\times 10^{23}$ cm$^2$ s$^{-1}$ would
correspond to turbulent transport at $1$ km s$^{-1}$ over $1$ pc. 

Clearly, the diffusivities are not constant, which agrees with the findings of 
\citet{DAM2002,DAM2003} and \citet{KLL2003}. In a fully turbulent medium - whose 
energy scale distribution follows at least qualitatively a turbulent spectrum -- 
the scale separation between turbulence and quantity to be diffused does not exist.

\subsection{Cloud Brightness}\label{ss:cloudbright}
Turbulence considerably changes the density profiles of our model cloud, and
reduces the optical depth at the center of the cloud. Does turbulent transport
(\S\ref{ss:turbdiss}) and the exchange of warm diffuse and cold dense 
material (\S\ref{ss:radprof}) carve tunnels and holes in the cloud
through which radiation can enter? For an answer, we have to determine the
radiation field within the cloud. \citet{BZH2004} discussed this question
with the help of a spherical cloud inscribed in a periodic box of 
evolved self-gravitating MHD-turbulence \citep{HMK2001,HZM2001}. This 
allowed them to study the radiation field inside an evolved structured
cloud. Here, we are interested in the timescales on which the
radiation field changes.

\begin{figure}
  \includegraphics[width=\columnwidth]{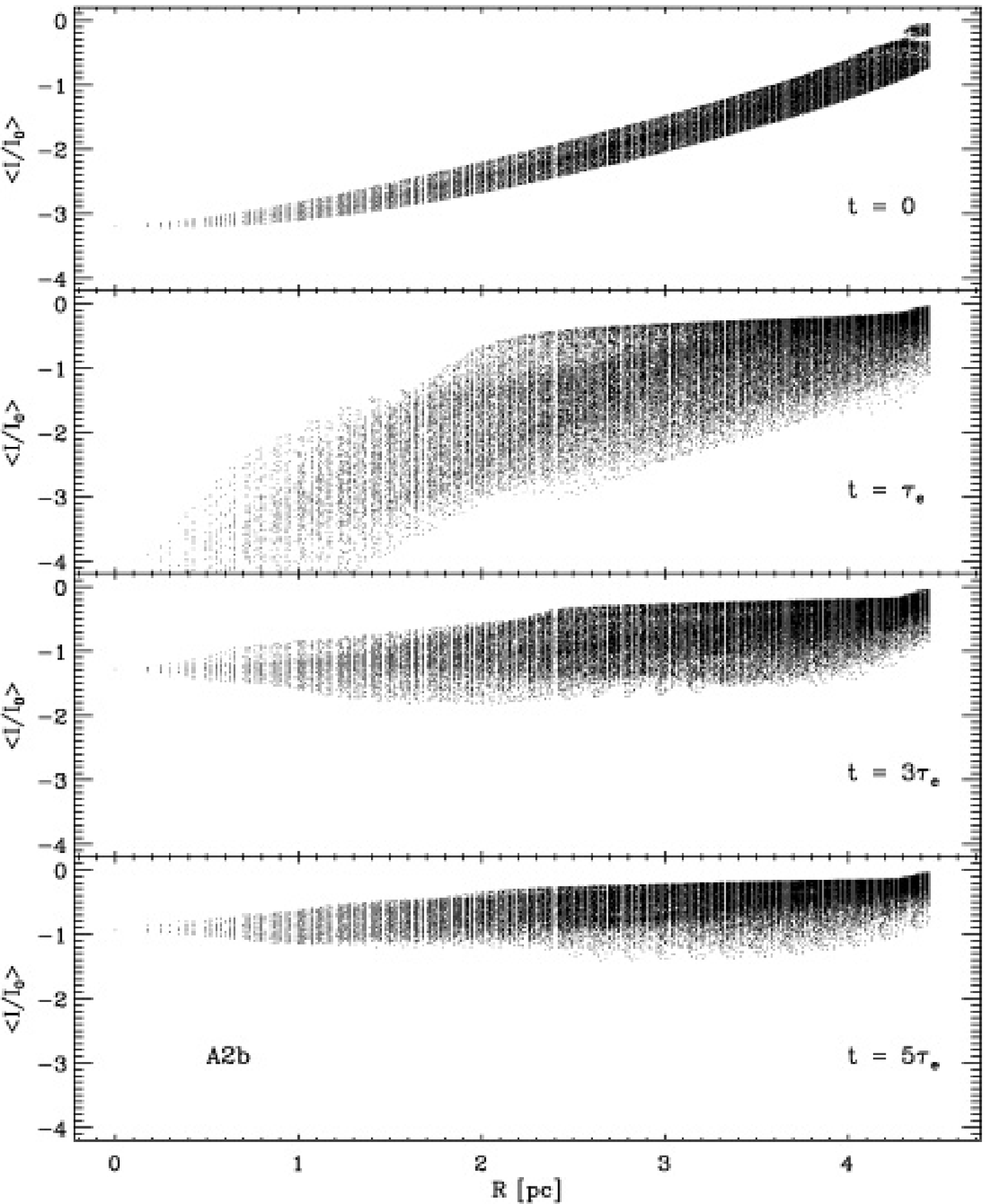}
  \caption{Scatterplots of relative brightness within cloud A2b, against radius, and
           for times as indicated in panels.
           \label{f:irrscatt}}
\end{figure}

To get a measure of the brightness inside the cloud, we determine the 
intensity of the radiation field at each point inside the cloud
(see \citealp{BZH2004}). Figure~\ref{f:irrscatt} shows a scatter plot
of the intensities inside the volume occupied by the original cloud 
against radius, for the four times indicated. 

The brightness is calculated for each point by measuring the incident
radiation for a given number of rays and averaging over the resulting
sky. The ray number is determined such that at the outer surface of the 
cloud each resolution element is hit by one ray. Note that while
we plot only points up to the original radius of the cloud, the radiative
transfer includes the whole domain, i.e. we do not lose ``dense'' material
outside $R=4.4$pc which could shadow the inner parts. For extinction,
we consider an effective extinction coefficient. Actual scattering is
not included. In that sense our brightness estimates are lower limits.

\begin{figure*}
  \includegraphics[width=\textwidth]{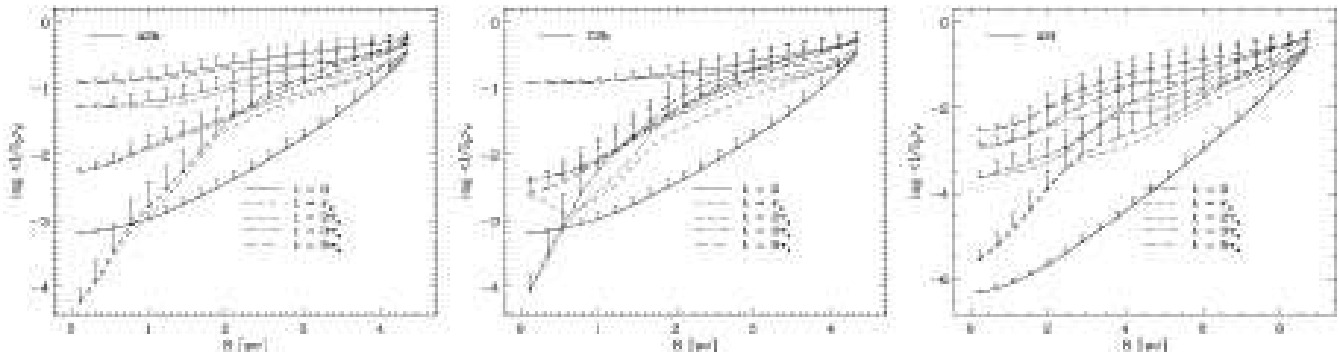}
  \caption{Radially binned intensities (see \S\ref{ss:cloudbright}) 
           for models A2b (left), C2b (center) and A2l (right). 
           Line styles denote the evolution time as indicated in
           the diagrams. Error bars represent errors on the mean. Thick lines
           are volume-averaged intensities, thin lines are mass-averaged intensities.
           \label{f:intensities}}
\end{figure*}

The radially binned intensities resulting from plots like Figure~\ref{f:irrscatt}
for models A2b, C2b and A2l 
are shown in Figure~\ref{f:intensities}.  Thick lines denote spherical
volume averages at constant $R$, and thin lines spherical mass averages.
Without self-gravity, the density contrasts in the cold material are
small, so that volume and mass averages do not differ widely. 
Clearly, Figure~\ref{f:intensities} mirrors the effect observed in Figure~\ref{f:irrscatt}:
the cloud gets ``bright'' within a few dynamical times, i.e. the turbulence
opens holes. Even for the larger cloud (A2l), the relative intensity 
does not drop below $10^{-3}$ anywhere in the cloud.

\begin{figure}
  \includegraphics[width=\columnwidth]{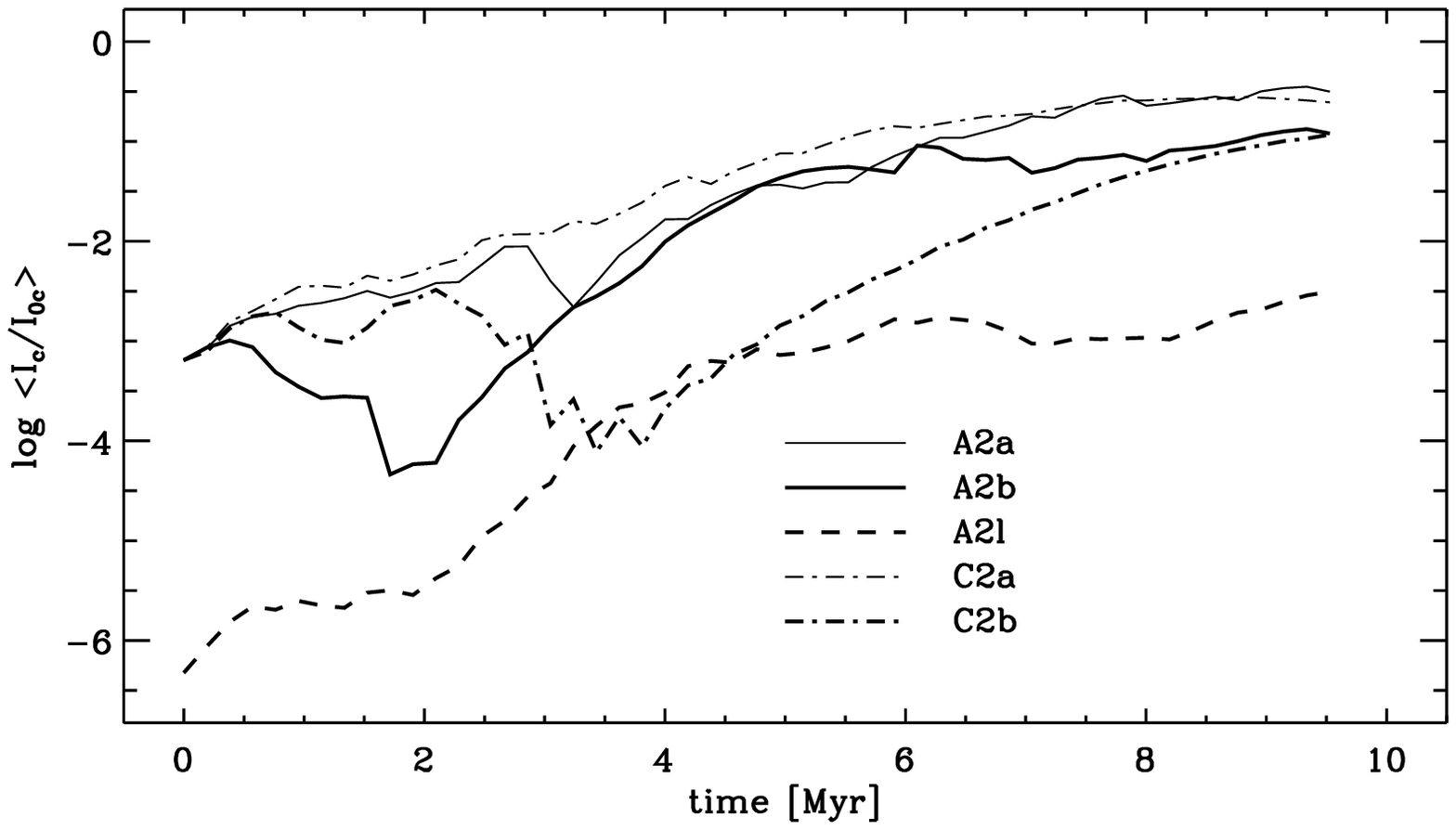}
  \caption{Central relative intensity against time for models A2a/b, C2a/b and
           A2l. Because of the larger cloud radius, model A2l stays at lower
           intensities. 
           \label{f:centertime}}
\end{figure}

At late times, even the innermost regions receive more than a 
tenth of the incident radiation (models A2a/b, C2a/b, Fig.~\ref{f:centertime}): 
turbulence (indirectly) lights up the cloud. The intensity at the center of the 
cloud increases by at least $2$ orders of magnitude. The central intensity for
the models at lower resolution (thin lines, A2a and C2a) grows more
smoothly than the intensity at higher resolution, and it reaches a slightly higher value
at the end of the simulation. These are both resolution effects: the larger scatter
comes from a more structured velocity field, and the larger central intensity
is a consequence of fewer available grid cells along the line of sight.
Still, for model A2l -- with its initial cloud radius twice the size of those
in models A2a/b --, the central intensity increases by four orders of magnitude
(compared to two for models A2a/b and C2a/b). Thus, we do not expect higher resolution to
lead to progressively smaller central intensities. 

\begin{figure*}
  \includegraphics[width=\textwidth]{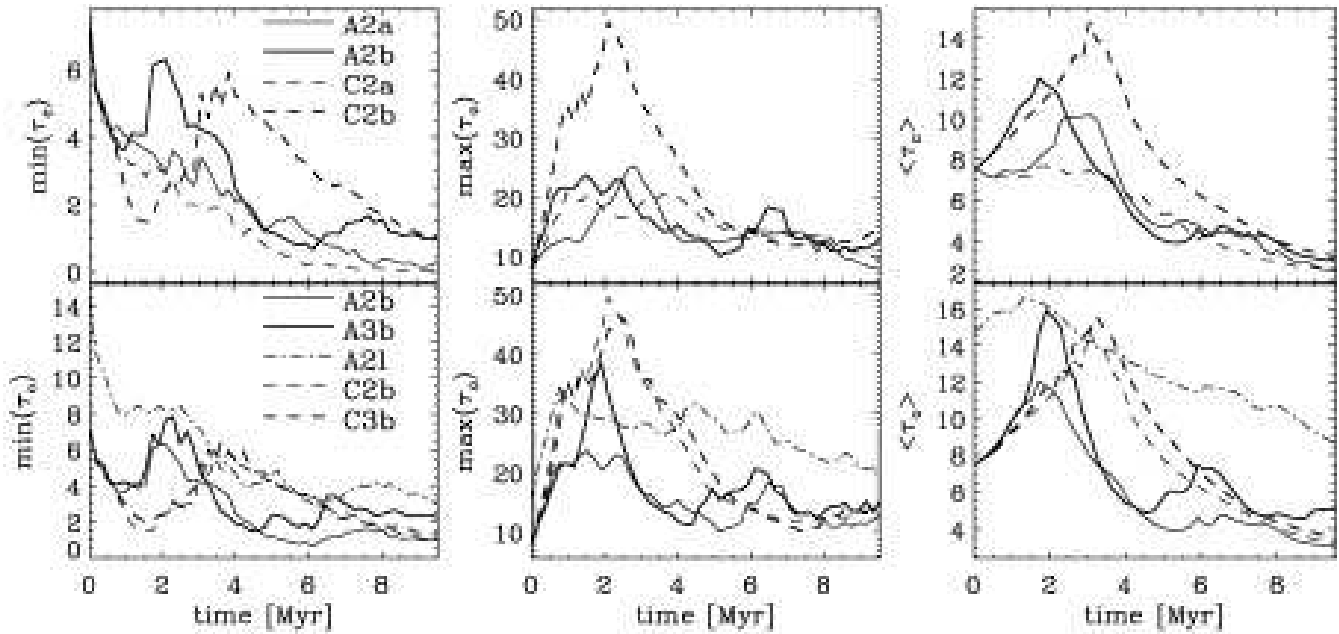}
  \caption{Optical depth at the center of the cloud against time.
           {\em Left column:} Minimum optical depth. {\em Center:} Maximum
           optical depth. {\em Right:} Average optical depth.
           {\em Top row:} Models at resolution $128^3$ (thin lines) and 
           $256^3$ (thick lines). 
           {\em Bottom row:} Parameter study. Thin lines denote models
           with an initial rms velocity of Mach 2, while thick lines 
           stand for those with Mach 3.
           \label{f:optdep}}
\end{figure*}

The central intensities discussed so far are derived from averaging the incoming
radiation over the whole ``sky'' as seen from the cloud center. If the turbulence
digs tunnels in the cloud through which the irradiation can enter deep into
the cloud, then this should be mirrored in the minimum (and maximum) optical
depth at the center of the cloud (Fig.~\ref{f:optdep}). The extrema of the optical
depths were taken over the whole sky as seen at the center of the cloud. The mean
optical depth corresponds to the center intensities discussed above (note however
that in order to derive the center intensities, the {\em intensities} are averaged 
over the sky, not the optical depths). 

As with the intensities (Fig.~\ref{f:centertime}), the lower resolution runs
lead to smaller average optical depths (Fig.~\ref{f:optdep}, top right). 
This is mostly a consequence of the lower minimum optical depths (top left of same
figure) and of the dominance of large-scale motions. Thus, the models run at
$128^3$ grid cells are not fully resolved, but should be used for demonstration purposes
only. The strong spike for model C2b is a direct effect of the radiative cooling:
compressed regions during the early turbulent evolution can reach much higher densities
than possible for the adiabatic case.
Although the minimum and maximum optical depth
starts out at the same value, the maximum optical depth increases due to local compressions. 
Of course, the directions of minimum and maximum optical depth change with time.
Models with different initial Mach number vary less than those with same Mach number but run
at different resolution. Since the turbulence in our models is decaying, this is not
surprising: most of the energy is lost in the early stages of the evolution. 

Summarizing Figure~\ref{f:optdep}, the minimum optical depths drop by a factor of $4$ to $8$
(depending on the model), while the maximum optical depths increase by a factor of $2$. Thus,
while most of the brightening of the cloud stems from turbulent transport of dense material
(see \S\ref{ss:dynphas} and \ref{ss:radprof}), compression and corresponding evacuation 
contributes as well. Finally, the average 
optical depths drop by a factor of $4$ or more. 

%
%
\section{Summary}\label{s:summary}

Motivated by the highly dynamical nature of molecular clouds
and the question of how long a cold cloud can survive as a 
well-defined entity in a turbulent environment,
we investigated the evolution of a
cold cloud in a warm, turbulent medium, assessing the efficiency
of the mixing by measuring the brightness inside the cloud. 
We showed that the properties of the initially solid cloud
change thoroughly within a few dynamical timescales.

While the column density maps can mislead, the 
brightness distribution inside the cloud clearly demonstrates that within
a few dynamical times, the cloud becomes completely porous. Specifically, the
cloud radius doubles within $5$ dynamical times, while the filling factor
of the cold gas drops to less than $50$\% at the original cloud radius within one 
dynamical time. Cold material is bodily transported and exchanged with warm 
material (Figs.~\ref{f:bulkmeas} and \ref{f:temphist}).
The turbulent diffusivity $\lambda_e$ (eq.~[\ref{e:diffdens}], [\ref{e:diffrms}] and 
Fig.~\ref{f:difftime})
is consistent with a fiducial number of $\lambda_e\approx 10^{23}$ cm$^2$ s${-1}$, 
corresponding to turbulent velocities and length scales of approximately $1$km s$^{-1}$ and 
$1$ pc. The diffusivities are not constant with time.

The brightness estimates are lower limits,
since we do not include scattering in the radiative transfer.
Gravity might change the results by leading to additional
fragmentation and thus to a growing discrepancy between volume-weighted 
and mass-weighted intensity \citep{BZH2004}. However, volumewise, the 
cloud would get even brighter this way due to the smaller filling factor 
of the dense gas. 
The effects of H$_2$ (re-)formation remain to be discussed in subsequent
models. If the turbulence were continually driven, we would expect an
even faster dispersal. In that sense also, the presented timescales are
only upper limits. 

The high radiation field within the cloud could strongly affect the 
chemistry and the dynamical state of the cloud \citep{DBC1991}, leading
to additional heating, H$_2$-destruction, and thus faster cloud disruption. 

\section*{Acknowledgements}
We thank the referee for a critical, very detailed and constructive report
which helped in reducing the number of faults and flaws in the manuscript.
The idea for the present study arose during the IAU Symposium 227, 
``Massive Star Birth: A Crossroads of Astrophysics''.
We enjoyed discussions with C.~F.~McKee about molecular cloud lifetimes
and turbulent dispersal, with K.~Menten about small-scale structure in 
molecular clouds, and with E.~Zweibel about turbulent transport.
Computations were performed at the NCSA (AST040026), and on the 
SGI-Altix at the USM, built and maintained by M.~Wetzstein and 
R.~Gabler. F.H. enjoyed the hospitality of CRAL both at the ENS Lyon 
and the Observatoire de Lyon during several visits. 

%
%

\bsp

\label{lastpage}

\end{document}